\documentclass[a4paper]{spie}  
\usepackage[]{graphicx}
\usepackage{color}
\usepackage{referj}
\title{Summary of the DUNE Mission Concept}

\author{A. Refregier\supit{a}, M. Douspis\supit{b} \& the DUNE collaboration\supit{c}
\skiplinehalf
\supit{a}SAp CEA Saclay, F-91191 Gif sur Yvette, France\\
\supit{b}IAS CNRS, b\^at. 121, Universit\'e Paris-Sud, F-91405 Orsay, France}

\authorinfo{\supit{c}{\bf DUNE collaboration:} {\bf Co-investigators:} A. Refregier ({\bf PI}, CEA Saclay) 
M. Douspis (IAS Orsay) Y. Mellier (IAP Paris) 
B. Milliard (LAM Marseille)
P. Schneider (U. Bonn) H.-W. Rix (MPIA) 
R. Bender (MPE Garching) F. Eisenhauer (MPE Garching) R. Scaramella (INAF-OARM)
% ...
\\ {\bf Authors'
  contacts:} refregier@cea.fr, marian.douspis@ias.u-psud.fr}

\begin{document} 
 \maketitle 

\begin{abstract}
The Dark UNiverse Explorer (DUNE) is a wide-field
imaging mission concept whose primary goal is the study of dark
energy and dark matter with unprecedented precision. To this end, DUNE
is optimised for weak gravitational lensing, and also uses complementary cosmological probes, such as baryonic oscillations, the integrated Sachs-Wolf effect, and cluster counts. Immediate additional goals concern the evolution of galaxies, to be studied with
groundbreaking statistics, the detailed structure of the Milky
Way and nearby galaxies, and the demographics of Earth-mass
planets. DUNE is a medium class mission consisting of a 1.2m telescope
designed to carry out an all-sky survey in one visible and three NIR
bands (1deg$^2$ field-of-view) which will form a unique legacy for
astronomy.  DUNE has been selected jointly with SPACE for an ESA Assessment
phase which has led to the Euclid merged mission concept.
\end{abstract}




\vspace*{3cm}

\section{INTRODUCTION}

\label{intro}
Dark energy and dark matter comprise the bulk of the mass-energy
budget of the Universe and pose some of the most fundamental questions
in physics.  The Dark UNiverse Explorer\footnote{{\tt
   http://www.dune-mission.net}} (DUNE) is a wide field mission
 concept designed to study these dark cosmic components with
 unprecedented precision. To do so, DUNE will use weak gravitational
 lensing along with other cosmological probes. In these proceedings,
 we give a brief summary of the DUNE mission concept which was
 recently proposed to ESA's Cosmic Vision programme. A description of
 the focal plane instrumentation can be found in an adjoining
 paper\cite{jeffmark} and a more detailed description of the ESA
 proposal is provided in a previous publication\cite{ref08}. An
 earlier and simpler version of DUNE was described in previous SPIE
 Proceedings\cite{gr06,ref06}.

Gravitational deflection of light by intervening dark matter
concentrations causes the images of background galaxies to acquire an
additional ellipticity of order of a percent, which is correlated over
scales of tens of arcminutes.  Utilisation of this cosmological probe
relies on the measurement of image shapes and redshifts for several
billion galaxies, both requiring space observations for PSF stablity
and photometric measurements over a wide wavelength range in the
visible and near-IR (NIR).

Furthermore, in order to break as many degeneracies as possible, and
to provide independent constraints, complementary approaches should be
used. DUNE has thus been designed to provide three additional
cosmological probes : Baryon Acoustic Oscillations (BAO), the
Integrated Sachs-Wolfe effect (ISW), and galaxy Cluster Counts
(CC). It is therefore a unique mission to probe the dark
 Universe in different independent ways.  DUNE will tackle the
following questions: What are the dynamics of dark energy?  What are
the physical characteristics of the dark matter?  What are the seeds
of structure formation and how did structure grow?  Is Einstein's
theory of General Relativity the correct theory of gravity?

DUNE will combine its unique space-borne observations with existing
and planned ground-based surveys, and hence greatly increases the
science return of the mission while limiting costs and risks. The
panoramic visible and NIR surveys required by DUNE's primary science
goals will afford unequalled sensitivity and survey area for further
studies.  Additional surveys at low galactic latitudes and in deep
patches of the sky will open new scientific windows. DUNE will explore
the nature of Dark Matter by measuring precisely the sum of the
neutrino masses and by testing the Cold Dark Matter paradigm. It will
test the validity of Einstein's theory of gravity. In addition, DUNE
will investigate how galaxies form, survey all Milky-Way-like galaxies
in the 2$\pi$ extra-galactic sky out to $z \sim 1$ and detect
thousands of galaxies and AGN at $6<z<12$. It will provide a detailed
visible/NIR map of the Milky Way and nearby galaxies and provide a
statistical census of exoplanets with masses above 0.1 Earth mass and
orbits greater than 0.5 AU.

Furthermore, DUNE surveys will provide a unique all-sky map in
 the visible and NIR at unprecedented spatial resolution for such
large scale of 0.25 and 0.3 arcesec respectively. It will thus
complement other space missions.

The DUNE mission will allow the investigation of a broad range of
astrophysics and fundamental physics. These and the corresponding
surveys are described in present and following section.  The last sections
describe the mission profile, payload instrument and the current
status of the mission concept.

\subsection{The Dark Universe}
\subsubsection{The nature of Dark Energy}

A variety of independent observations overwhelmingly indicate that the
cosmological expansion began to accelerate when the Universe was
around half of its present age. Presuming the correctness of General
Relativity this requires a new energy component known as Dark Energy.
The simplest case would be Einstein Cosmological Constant ($\Lambda$),
in which the dark energy density would be exactly homogeneous and
independent of time. However, such an interpretation conflicts with
predictions of vacuum energy from Particle Physics by 120 orders of
magnitude. For this reason, cosmologists are motivated to consider
models of a dynamical dark energy, or even to contemplate
modifications to General Relativity.

DUNE will deduce the expansion history using the distance-redshift
relation ($D(z)$) and growth of structure. It can thus probe the
nature and properties evolution of dark energy in two independent
ways.  A single accurate technique can rule out many of the suggested
members of the family of dark energy models, but it cannot test the
fundamental assumptions about gravity theory. If General Relativity is
correct, then either $D(z)$ or the growth of structure can determine
the expansion history. In more radical models that violate General
Relativity, however, this equivalence between $D(z)$ and growth of
structure does not apply.  For this purpose, DUNE will use a
combination of the following cosmological probes.

{\it Weak Lensing - A Dark Universe Probe:} As light from galaxies
travels towards us, its path is deflected by the intervening mass
density distribution, causing the shapes of these galaxies to appear
distorted by a few percent. The weak lensing method measures this
distortion by correlating the shapes of background galaxies to probe
the density field of the Universe. By dividing galaxies into redshift
(or distance) bins, we can examine the growth of structure (as a
function of redshift) and make three-dimensional maps of the dark
matter. As the evolution of the growth of structure is different in
different scenarios of Dark Energy and dark matter, we have access to
the nature of these dark components.  An accurate lensing survey,
therefore, requires precise measurements of the shapes of galaxies as
well as information about their redshifts. High-resolution images of
large portions of the sky are required, with low levels of systematic
errors that can be achieved only via observations from a thermally
stable satellite in space. Analyses of the dark energy require precise
measurements of both the cosmic expansion history and the growth of
structure. Weak lensing stands apart from all other available methods
because it is able to make accurate measurements of both
effects\cite{ama07,ama07b}.  Given this, the optimal dark energy
mission (and dark sector mission) is one that is centred on weak
gravitational lensing and is complemented by other dark energy probes.

{\it Baryon Acoustic Oscillations (BAO) -- An Expansion History
 Probe:} The scale of the acoustic oscillations caused by the
coupling between radiation and baryons in the early Universe can be
used as a 'standard ruler' to determine the distance-redshift
relation. Using DUNE, we can perform BAO measurements using
photometric redshifts yielding the three-dimensional positions of a
large sample of galaxies. All-sky coverage in the NIR enabled by DUNE,
but impossible from the ground, is essential to reach the necessary
photometric redshift accuracy for this BAO survey.

{\it Cluster Counts (CC) -- A Growth of Structure Probe:} Counts of
the abundance of galaxy clusters (the most massive bound objects in
the Universe) as a function of redshift are a powerful probe of the
growth of structure. There are several ways to exploit the DUNE
large-area survey for cluster detection: weak lensing, strong lensing,
optical richness, and cross-correlation with X-ray or Sunyaev-Zeldovich
surveys (eg. Planck, E-Rosita, etc).

{\it Integrated Sachs-Wolfe (ISW) Effect -- A Higher Redshift Probe:}
The ISW effect is the change in CMB photon energy as it passes through
a changing potential well.  Its presence indicates either space
curvature, a dark energy component or a modification to gravity. The
ISW effect is measured by cross-correlating the CMB with a foreground
density field covering the entire extra-galactic sky, as measured by
DUNE. The presence and shape of the spectrum of this cross-correlation
will thus inform us about the nature of dark energy. Because it is a
local probe of structure growth, ISW will place complementary
constraints on dark energy, at higher redshifts, relative to the other
probes\cite{dous08}.

\begin{figure}
\begin{center}
 \begin{tabular}{c}
\includegraphics[width=0.4\textwidth,angle=0]{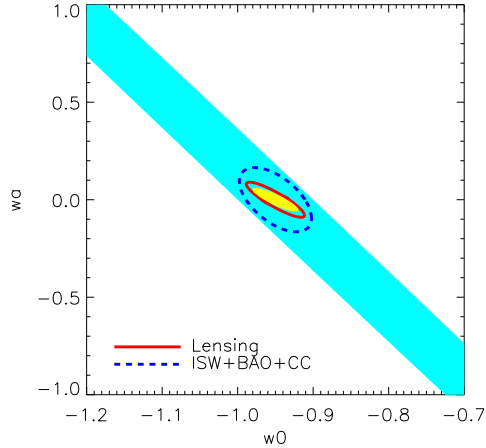}
\end{tabular}
\end{center}
\caption{Expected errors on the dark energy equation of state
 parameters (in a scenario where $w(a) = w_0+(1-a) w_a$) for the four
 probes used by DUNE. The light blue band indicates the expected
 errors from Planck. By combining BAO, CC and ISW we begin to reach
 similar accuracies to that of the lensing survey. This combination
 thus gives DUNE two independent and comparable measures of dark
 energy. }
\label{figfom}
\end{figure}

The combination of all these probes will provide strong constraints
on the properties and evolution of dark energy (see Fig.~\ref{figfom}).

\subsubsection{Properties of Dark Matter}

Besides dark energy, one major component of the concordance model of
cosmology is dark matter ($\sim90$\% of the matter in the Universe,
and $\sim 25$\% of the total energy). The standard assumption is that
the dark matter particle(s) is cold and non-collisional (CDM). Its
nature \emph{may} well be revealed by experiments such as the Large
Hadron Collider (LHC) at CERN, but its physical properties may prove
to be harder to pin down without astronomical input. One way of
testing this is to study the amount of substructure in dark matter
halos on scales 1-100'', which can be done using high order galaxy
shape measurements and strong lensing with DUNE. Weak lensing measures
with DUNE can constrain the total neutrino mass and number of neutrino
species through observations of damping of the matter power spectrum
on small scales. Combining DUNE measurements with Planck data would
reduce the uncertainty on the sum of neutrino masses to 0.04eV, and
should therefore measure the neutrino mass\cite{kit08}.

\subsubsection{The Seeds of Structure Formation}

It is widely believed that cosmic structures originated during
inflation from vacuum fluctuations in primordial quantum fields
stretched to cosmic scales. In the most basic inflation models, the
power spectrum of these fluctuations is predicted to be close to
scale-invariant.

As the Universe evolved, these initial fluctuations grew.  CMB
measurements probe their imprint on the radiation density at $z \sim
1100$. Density fluctuations continued to grow into the structures we
see today. Combined with Planck, the DUNE weak lensing observations and
auto-correlation function of galaxies  will provide a measurement
of the shape of the primordial spectrum with unprecedent accuracy
($1\%$ on $\sigma_8$ and $n_s$).

\subsubsection{Probing Einstein's Gravity}

Various modifications to gravity on large scales (e.g. by extra
dimensions, superstrings, non-minimal couplings or additional fields)
have been suggested to avoid the need to invoke dark matter and dark
energy. The weak lensing measurements of DUNE will be used to test the
underlying theory of gravity, using the fact that modified gravity
theories typically alter the relation between geometrical
 measure and the growth of structure.  DUNE will be able to measure
the growth factor exponent $\gamma$ (signature of the deviation from Einstein gravity\cite{amend08}) with a precision of 2\%.

\subsection{Tracking the Formation of Galaxies and AGNs}
In order to disentangle the complex processes involved in galaxy
formation, we need high-resolution near-IR space-based images to study
galaxy morphology and large area for crucial, rare events, such as the
merger rate of very massive galaxies; DUNE will provide this to $z
\sim 1.5$.

Using DUNE's weak lensing maps, we will study the relationship between
galaxy mass and light, the bias, by mapping the total mass density and
stellar mass, luminosity, morphological type of glaxies.  While at
present only a few massive clusters at $z>1$ are known, DUNE will find
hundreds of Virgo-cluster-mass objects at $z>2$, and several thousand
clusters of M=$1-2 \times 10^{13}$Mo; the latter are the likely
environments in which the peak of QSO activity at $z\sim2$ takes
place, and will hold the empirical key to understanding the peak period of
QSO activity.

Using the Lyman-dropout technique in the near-IR, a deep survey
(DUNE-MD) will be able to detect the most luminous objects in the
early Universe ($z>6$): $\sim 10^4$ star-forming galaxies at $z\sim8$
and up to $10^3$ at $z\sim10$, for SFRs $>30-100$Mo/yr. It will also
be able to detect significant numbers of high-$z$ quasars: up to
$10^4$ at $z\sim7$, and $10^3$ at $z\sim9$. By applying the
Gunn-Peterson test to this large statistically relevant sample of
objects we will put stringent constraints on the end of the period of
reionisation of the Universe.

DUNE will also detect a very large number of strong lensing
 systems: about $10^5$ galaxy-galaxy lenses, $10^3$ galaxy-quasar
lenses and 5000 strong lensing arcs in clusters\cite{meneg07}. It is
also estimated that several tens of galaxy-galaxy lenses will be
double Einstein rings\cite{gav08}, which are powerful probes
of the cosmological model as they simultaneously probe several
redshifts.

In addition, during the course of the DUNE-MD survey (over 6 months),
we expect to detect $\sim 3000$ Type Ia Supernovae with
redshifts up to $z\sim0.6$. This will yield a measurement of the SN
rate with unprecedented accuracy, thus providing information on their
progenitors. This survey will also discover a comparable number of
Core Collapse SNe (Types II and Ib/c), out to $z\sim0.3$, whose rate
provides an independent measurement of the star formation history.

\subsection{Studying the Milky Way}

DUNE will also leas to breakthroughs in Galactic astronomy. The
extragalactic survey, DASS-EX, complemented by the shallower survey of
the Galactic plane (DASS-G with $|b|<30$) will provide all-sky high
resolution (0.23'' in the wide red band, and 0.4'' in YJH) deep
imaging of the stellar content of the Galaxy, allowing the deepest
detailed structural studies of the thin and thick disk components, the
bulge/bar, and the Galactic halo (including halo stars in nearby
galaxies such as M31 and M33) in bands which are relatively
insensitive to dust in the Milky Way.

DUNE will be little affected by extinction and will supersede all of
the ongoing surveys in terms of angular resolution and sensitivity
(photometric depth and low background).  DUNE will thus enable the
most comprehensive stellar census of late-type dwarfs and giants,
brown dwarfs, He-rich white dwarfs, along with detailed structural
studies, tidal streams and merger fragments.  DUNE's sensitivity will
also open up a new discovery space for rare stellar and
low-temperature objects via its H-band imaging.  Studying the Galactic
disk components requires the combination of spatial resolution
(crowding) and dust-penetration (H-band) that only DUNE can
deliver. It will also yield the most detailed and sensitive survey of
structure and substructure in nearby galaxies especially of
their outer boundaries, thus constraining merger and accretion
histories.

\subsection{Search for Extra-Solar Planets}

Using the microlensing effect, DUNE can provide a statistical census
of exoplanets in the Galaxy with masses over $0.1 M_\oplus$ from
orbits of 0.5 AU from their parent star to free-floating objects. This
includes analogues to all the solar system's planets except for
Mercury, as well as most planets predicted by planet formation theory.
Microlensing is the temporary magnification of a galactic bulge source
star by the gravitational potential of an intervening lens star
passing near the line of sight. A planet orbiting the lens star, will
have an altered magnification, showing a brief flash or a dip in the
observed light curve.  Because of atmospheric seeing, and poor duty
cycle even using networks, ground-based microlensing surveys are only
able to detect a few to 15 $M_\oplus$ planets in the vicinity of the
Einstein ring radius (2-3 AU). A dedicated survey (DUNE-ML), using the
high angular resolution of DUNE, and the uninterrupted visibility and
NIR sensitivity afforded by space observations will provide detections
of microlensing events using as sources G and K bulge dwarfs stars and
therefore can detect planets down to $0.1-1 M_\odot$ from orbits of
0.5 AU.  Moreover, there will be a very large number of transiting hot
Jupiters detected towards the galactic bulge as a free ancillary
science.

\section{Surveys}\label{toolssurveys}

To achieve the scientific goals described above, DUNE
will perform four surveys detailed in the following.

\begin{figure}
\begin{center}
 \begin{tabular}{c}
\includegraphics[width = 0.4\textwidth, angle=90]{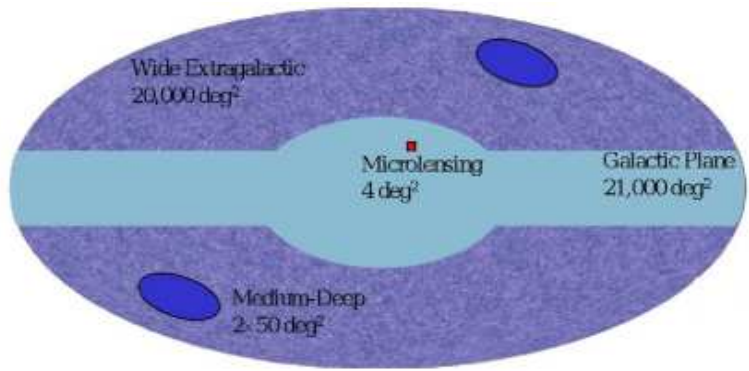}
\end{tabular}
\end{center}
\caption{Schematic view of the surveys performed by DUNE}
\end{figure}

\subsection{Wide Extragalactic Survey: DASS-EX }

To measure dark energy to the required precision we propose a high
precision weak lensing survey with a large area to provide large
statistics and the control of systematic errors. DUNE will make
measurements over the entire extra-galactic sky (20000 deg$^2$) to a
depth which yields 40 gal/arcmin$^2$ useful for lensing with a median
redshift $z_m \simeq 0.9$. This will be achieved with a survey (DASS-EX) that
has AB-magnitude limit of 24.5 (10$\sigma$ extended source) in a broad
red visible filter (R+I+Z). Based on the fact that DUNE will focus on
observations that cannot be obtained from the ground, the wide survey
relies on two unique factors that are enabled by space: image quality
in the visible and NIR photometry.  Central to shape measurements for
weak lensing the PSF of DUNE needs to be sampled better than 2.3
pixels per FWHM, to be constant over 50 stars around each galaxy
(within a tolerance of $\sim 0.1\%$ in shape parameters), and to have
a calibratable wavelength dependence\cite{paul07}.  Accurate
measurement of the redshift of distant galaxies ($z \sim 1$) requires
photometry in the NIR where galaxies have a distinctive feature (the
4000$A$ break).  Deep NIR photometry requires space observations. The
wide survey will provide NIR photometry in Y, J and H down to AB-magnitude
of 24 (5$\sigma$ Point Source) providing an ideal synergy for ground based survey
complement(\cite{abda07}), as recommended by the ESO/ESA Working Group
on Fundamental Cosmology\cite{peack06}.


\subsection{DUNE-MD}

We propose to allot six months to a medium deep survey (DUNE-MD) with
an area of 100 deg$^2$ to magnitudes of 26 in Y, J and H, located at
the North and South ecliptic poles. This survey can be used to
calibrate DUNE during the mission, by constructing it from a stack of
$>30$ sub-images to achieve the required depths. The PSF properties
would be the same as the DASS-EX survey.

\subsection{DASS-G}

DUNE will also perform a wide galactic survey (DASS-G) with shorter
exposures on the remaining 20000 deg$^2$. The limit AB-magnitude would
be 23.8 in visible and 22 in NIR.

\subsection{DUNE-ML}

DUNE will also perform a microlensing survey (DUNE-ML) with short
exposures over 4 deg$^2$ in the galactic bulge.  Besides the DASS-EX
and DUNE-MD, this survey needs low levels of stray light.

\section{Mission Profile and Payload instrument} 

The mission design of DUNE is driven by the need for the stability of
the PSF and large sky coverage. PSF stability puts stringent
requirements on pointing and thermal stability during the observation
time. The 20,000 deg$^2$ survey (DASS-EX) demands high operational
efficiency, which can be achieved using a drift scanning mode (or Time
Delay Integration, TDI, mode) for the CCDs in the visible focal
plane. TDI mode necessitates the use of a counter-scanning mirror to
stabilize the image in the NIR focal plane channel. The baseline orbit
for DUNE is a Geosynchronous Earth orbit (GEO) with an Highly Elliptical
Orbit (HEO) as a possible alternative.

\begin{figure}
\begin{center}
\begin{tabular}{c}
\includegraphics[width = 0.5\textwidth, angle=0]{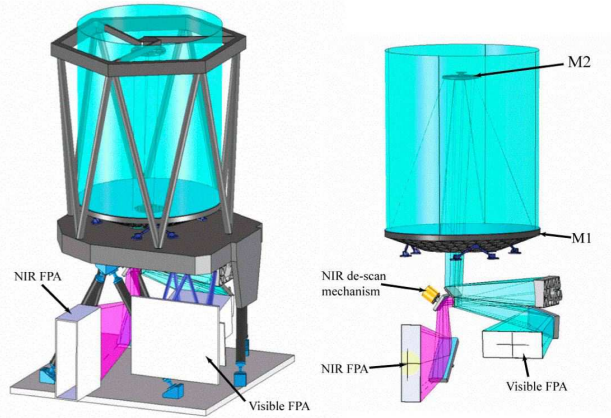}
\end{tabular}
\end{center}
\caption{Telescope and focal planes designs}
\label{fig:instrument}
\end{figure}

The telescope is a passively cooled 1.2m diameter Korsch-like $f/20$
three-mirror telescope. with two focal planes, visible and NIR
covering together 1 deg$^2$ (see Figure \ref{fig:instrument}). After the first two mirrors, the optical bundle is
folded just after passing the primary mirror (M1) to reach the
off-axis tertiary mirror. A dichroic element located near the exit
pupil of the system provides the spectral separation of the visible
and NIR channels.

The visible Focal Plane Array (VFP) consists of 36 large format
red-sensitive CCDs, arranged in a $9 \times 4$ array.  Four additional CCDs
dedicated to the attitude control (AOCS) measurements are located at
the edge of the array. All CCDs are 4096 pixel red-enhanced e2v
CCD203-82 devices with square 12 $\mu$m pixels.  The physical size of
the array is 466x233 mm corresponding to $1.09\deg \times 0.52
\deg$. Each pixel is 0.102 arcsec, so that the PSF is well sampled in
each direction over approximately 2.2 pixels, including all
contributions.  The VFP operates in the red band from 550-920nm.

The VFP will be used by the spacecraft in a closed-loop system to
ensure that the scan rate and TDI clocking are synchronised. The two
pairs of AOCS CCDs provide two speed measurements on relatively bright
stars (V $\sim 22-23$). The DUNE VFP is largely a self-calibrating
instrument. For the shape measurements, stars of the appropriate
magnitude will allow the PSF to be monitored for each CCD including
the effects of optical distortion and detector alignment.
Radiation-induced charge transfer inefficiency will modify the PSF and
will also be calibrated through in-orbit self-calibration.

The NIR FPA consists of a $5 \times 12$ mosaic of 60 Hawaii 2RG detector
arrays from Teledyne, NIR bandpass filters for the wavelength bands Y,
J, and H, the mechanical support structure, and the detector readout
and signal processing electronics. The NIR FPA is operated at a
maximum temperature of 140 K for low dark current of 0.02$e^-$/s. Each
array has 2048 x 2048 square pixels of 18 $\mu$m size resulting in a
$0.15 \times 0.15$ arcsec$^2$ field of view (FOV) per pixel.  The mosaic has
a physical size of $482 \times 212$ mm, and covers a FOV of $1.04^\circ
\times 0.44^\circ$ or 0.46 square degrees.  The HgCdTe Hawaii 2RG
arrays are standard devices sensitive in the 0.8 to 1.7 $\mu$m
wavelength range.

As the spacecraft is scanning the sky, the image motion on the NIR FPA
is stabilised by a de-scanning mirror during the integration time of
300s or less per NIR detector. The total integration time of 1500 s
for the $0.4^\circ$ high field is split among five rows and 3
wavelengths bands along the scan direction.  The effective integration
times are 600 s in J and H, and 300 s in Y. For each array, the
readout control, A/D conversion of the video output, and transfer of
the digital data via a serial link is handled by the SIDECAR ASIC
developed for JWST. To achieve the limiting magnitudes defined by the
science requirements within these integration times, a minimum of 13
reads are required.  Data are processed in the dedicated unit located
in the service module.

The spacecraft platform architecture is fully based on well-proven and
existing technologies. The mechanical, propulsion, and Solar array
systems are reused from Venus Express (ESA) and Mars-Express. All the
AOCS, $\mu$-propulsion, Power control and DMS systems are reused from
GAIA. Finally, the science telemetry system is a direct reuse from the
PLEIADES (CNES) spacecraft.

\section{Conclusions}

\begin{table}
\begin{center}
\caption{DUNE Baseline summary} 
\label{baseline}
\begin{tabular}{|l|l|}
\hline
Science objectives & Cosmology and Dark Energy\\
&  Galaxy formation,   Extra-solar planets\\
\hline
Surveys &  20,000 deg$^2$ extragalactic 20,000 deg$^2$  galactic\\ 
& 100 deg$^2$ medium-deep, 4 deg$^2$ planet hunting\\
\hline
Requirements & 1 visible band (R+I+J) for high-precision\\
& shape measurements\\
&  3 NIR bands (Y, J, H) for photometry\\
\hline
Payload & 1.2m telescope, Visible \& NIR cameras \\
& with 0.5 deg$^2$ FOV each\\
\hline
Service module & Mars/Venus express, Gaia heritage \\
\hline
Spacecraft & 2013kg launch mass\\
\hline
Orbit & Geosynchronous\\
\hline
Launch & Soyuz S-T Fregat\\
\hline
Operations & 4 year mission\\
\hline
\end{tabular}

\end{center}
\end{table}

The DUNE mission concept can be seen as the next step in precision
cosmology. ESA's Planck 
mission will bring unprecedented
precision to the measurement of the high redshift Universe. This will
leave the dark energy dominated low redshift Universe as the next
frontier in high precision cosmology. Constraints from the radiation
perturbation in the high redshift CMB, probed by Planck, combined with
density perturbations at low redshifts, probed by DUNE, will form a
complete set for testing all sectors of the cosmological model. In
this respect, a DUNE+Planck programme can be seen as the next
significant step in testing, and thus challenging, the standard model
of cosmology. DUNE will offer high potential for ground-breaking
discoveries of new physics, from dark energy to dark matter, initial
conditions and the law of gravity.  DUNE will (i) measure both effects
of dark energy by using weak lensing as the central probe; (ii) place
this high precision measurement of dark energy within a broader
framework of high precision cosmology by constraining all sectors of
the standard cosmology model (dark matter, initial conditions and
Einstein gravity); (iii) through a collection of unique legacy surveys
be able to push the frontiers of galaxy evolution and the physics of
the local group; and finally (iv) be able to obtain information on
extrasolar planets, including Earth analogues. 

The DUNE concept has been recently proposed to ESA's Cosmic Vision
programme and has been selected jointly with SPACE\cite{space} for an ongoing ESA Assesment
Phase which has led to the merged {\it Euclid} mission concept.

\acknowledgments     

We thank CNES for support on an earlier version of the DUNE mission
and EADS/Astrium, Alcatel/Alenia Space, as well as Kayser/Threde for
their help in the preparation of the ESA proposal.


\bibliography{dune_spie}   
\bibliographystyle{spiebib}   

\end{document}